\documentclass[%
reprint,
superscriptaddress,
prl,
amsmath,amssymb,
aps,
]{revtex4-2}

\usepackage{color}
\usepackage{gensymb}
\usepackage{graphicx}
\usepackage{dcolumn}
\usepackage{bm}
\usepackage{hyperref}
\usepackage{epstopdf}
\usepackage{float}
\usepackage{listings}
\usepackage{times,mathptm,bm}
\usepackage{xcolor}
\usepackage[export]{adjustbox}
\usepackage[version=4]{mhchem}
\usepackage{siunitx}
\usepackage{enumitem}
\usepackage{booktabs}
\usepackage{bibunits}
\defaultbibliographystyle{apsrev4-2}
\defaultbibliography{biblio}

\begin{document}
\begin{bibunit}
\title{Active Reinforcement of Jammed Emulsions by Living Microswimmers}

\author{Marie Corpart}
\affiliation{Van der Waals-Zeeman Institute, Institute of Physics, University of Amsterdam, 1098 XH Amsterdam, Netherlands.}
\author{Hugo Le Roux}
\affiliation{Van der Waals-Zeeman Institute, Institute of Physics, University of Amsterdam, 1098 XH Amsterdam, Netherlands.}
\author{Daniel Bonn}
\affiliation{Van der Waals-Zeeman Institute, Institute of Physics, University of Amsterdam, 1098 XH Amsterdam, Netherlands.}
\author{Antoine Deblais}
\email{a.deblais@uva.nl}
\affiliation{Van der Waals-Zeeman Institute, Institute of Physics, University of Amsterdam, 1098 XH Amsterdam, Netherlands.}

\date{\today}

\begin{abstract}
We show that living microswimmers mechanically reinforce dense emulsions.
Castor-oil-in-water emulsions laden with the microalga \textit{Chlamydomonas reinhardtii}
are compared in three states: without algae, with immobilized algae, and with motile algae,
over a broad range of oil fractions spanning the jamming transition.
Oscillatory rheology reveals that motile algae systematically increase the yield stress,
by up to a factor of two, whereas immobilized cells at the same concentration
leave it essentially unchanged.
The reinforcement thus originates from activity rather than from the mere presence of inclusions.
Single-cell tracking shows that the droplet network confines the swimmers in pores
that shrink as the oil fraction increases, and confinement is known to amplify
the propulsion force of \textit{C.~reinhardtii}.
A simple estimate based on this confinement-enhanced force accounts
for the measured excess yield stress and indicates an effective, activity-induced
depletion-like attraction between the passive droplets.
These results identify a feedback loop: the microstructure confines the swimmers,
confinement amplifies the forces they exert, and these forces stiffen the microstructure.
Active emulsions thus emerge as a model platform for programming the mechanics
of disordered soft solids through activity.
\end{abstract}

\maketitle

\textit{Introduction.}--Jamming is a fundamental transition in soft matter,
where disordered packings of particles acquire rigidity once their packing fraction
$\phi$ exceeds a critical threshold $\phi_{\rm c}$.
In athermal, purely repulsive systems such as emulsions and foams,
this threshold coincides with random close packing (RCP):
above RCP the material supports a finite yield stress $\sigma_y$~\cite{Mason1996},
whereas below RCP it flows as a shear-thinning non-Newtonian fluid \cite{vanHecke2010,Bonn2017}.
Emulsions are a particularly clean platform to probe jamming:
they consist of deformable droplets with well-defined interfacial tension,
tunable volume fraction, and well-characterized rheology.
Classical studies have established how viscoelastic moduli and yield stress
scale near RCP and depend on droplet size, polydispersity,
and interdroplet interactions \cite{Mason1996,Bonn2017,Scheffold2013}.

Introducing attractive interactions between droplets, through depletion by surfactant micelles
or through surface chemistry, qualitatively changes this picture.
Attractive emulsions form loosely connected networks and flocs,
and therefore sustain a finite yield stress and solid-like response
even below $\phi_{\rm c}$, whereas above $\phi_{\rm c}$ their elasticity
becomes indistinguishable from that of repulsive packings~\cite{Datta2011}.
Attraction also alters the way these materials yield,
through bond breaking, shear localization, and slow structural relaxation
absent in the non-adhesive case~\cite{Becu2006,Irani2014}.
The rigid--fluid boundary is thus controlled not by packing density alone,
but jointly by the microstructure and by the interaction potential.

Whether--and how--\emph{active} inclusions displace this boundary is far less clear.
Numerical studies of dense active Brownian particles predict that self-propulsion
can shift the jamming point to higher volume fractions and modify the scaling
of diffusivities and yield stresses near the transition,
promoting unjamming at high propulsion speeds~\cite{Ni2013,Bi2016}.
Experiments point in the same direction for pusher-type swimmers:
bacterial suspensions display reduced viscosity, shear thinning,
and even a superfluid-like regime~\cite{Sokolov2009,Lopez2015,Martinez2020}.
Puller-type swimmers behave oppositely: the viscosity of
\textit{C.~reinhardtii} suspensions is markedly \emph{larger}
than that of dead cells at the same volume fraction~\cite{Rafai2010,Mussler2013}.
In parallel, active baths are known to mediate effective attractions between passive bodies
that closely resemble depletion forces \cite{Angelani2011,Grober2023},
and confinement strongly modifies both the trajectories and forces
of individual swimmers \cite{Bechinger2016,Boddeker2020}.
What has not been addressed experimentally is the regime where all these ingredients meet:
active particles of a size comparable to that of the passive constituents,
embedded in and caged by a jammed soft solid, and mechanically coupled to it.

Here we introduce attractive emulsions containing motile
\textit{Chlamydomonas reinhardtii} to probe directly how activity controls
the rheology of a droplet network.
Comparing emulsions without algae, with immobilized algae, and with motile algae
across oil fractions $\phi_{\rm oil}$ spanning below, at, and above $\phi_{\rm c}$,
and combining oscillatory rheometry with microscopic tracking of the algal motion,
we find that living inclusions systematically increase the yield stress $\sigma_y$
at all $\phi_{\rm oil}$ investigated, including below the classical random close packing threshold.
We further show that motility becomes progressively constrained at high oil fractions
because of caging by neighbouring droplets, which in turn amplifies the active forces
exerted by the confined algae on the network.

These results highlight the role of activity in jammed emulsions:
living motile inclusions reinforce the droplet network because confinement
increases the active stresses they generate.
This coupling between microscopic activity and macroscopic rigidity
provides a route to tune the rheology of dense emulsions through activity and confinement,
and suggests new strategies for designing active materials
with programmable mechanical properties.

\begin{figure}[t]
\centering
\includegraphics{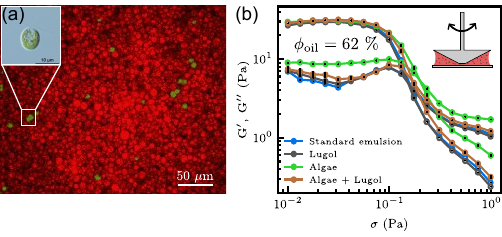}
\caption{\textbf{Active emulsions.}
(a) Reconstructed confocal image of an emulsion at $\phi_{\rm oil}=0.50$.
Oil droplets (red) and algae (green) are visualized by Nile Red and chlorophyll fluorescence, respectively.
(b) Storage modulus $G'$ (open symbols) and loss modulus $G''$ (filled symbols)
as a function of the applied shear stress, for emulsions at $\phi_{\rm oil}=0.62$
without algae, with motile algae, and with algae immobilized by Lugol's solution.}
\label{fig:ActiveEmulsionSnapshot_and_OscillationMeasurement}
\end{figure}

\begin{figure*}[t]
\centering
\includegraphics[width=0.8\textwidth]{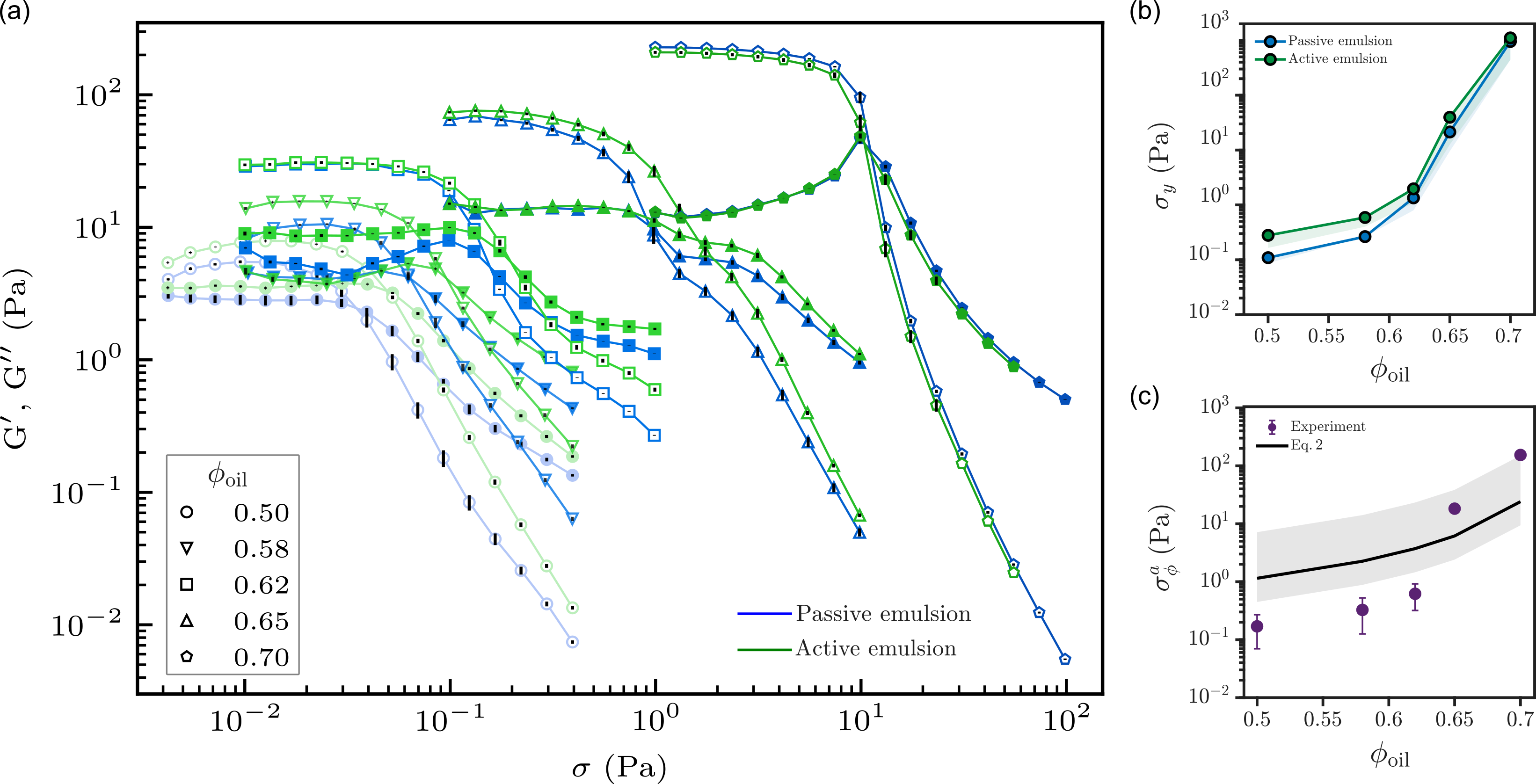}
\caption{\textbf{Rheological response of passive and active emulsions.}
(a) Storage $G'$ (open symbols) and loss $G''$ (filled symbols) moduli
as a function of shear stress for emulsions without (blue) and with (green) motile algae,
shown for selected oil fractions $\phi_{\rm oil}$ (marker shape and transparency indicate $\phi_{\rm oil}$).
(b) Yield stress $\sigma_y$ of passive (blue) and active (green) emulsions as a function of oil fraction.
(c) Active contribution to the yield stress, $\sigma_\phi^{\rm a}=\sigma_y^{\rm a}-\sigma_y^{\rm p}$,
as a function of oil fraction.
The solid line is the prediction of Eq.~(\ref{eq:sigma_phi}); the grey shaded region represents
the uncertainty on this prediction, dominated by the spread in droplet radius
and in the reported propulsion forces.
In (a) and (b), each point is the average over 1--5 independent emulsions,
with at least four oscillatory measurements per sample;
error bars denote the standard error of the mean.}
\label{fig:Results_ShearRheologyOscillation_All}
\end{figure*}

\textit{Experimental system.}--Our model system consists of castor oil-in-water emulsions
in which motile \textit{Chlamydomonas reinhardtii} cells are dispersed in the continuous phase.
Droplets are stabilized with the biocompatible surfactant Pluronic F-127
(CMC $=3.3$ g/L in TAP (Tris--Acetate--Phosphate) growth medium),
used at $c_{\rm surf}=10$ CMC, and the oil phase is labelled with Nile Red
for confocal imaging [red in Fig.~\ref{fig:ActiveEmulsionSnapshot_and_OscillationMeasurement}(a)],
which gives direct access to droplet size and network arrangement.
Emulsions are prepared in two steps: a high-shear emulsification (Silverson L5M-A)
produces a concentrated mother emulsion (internal fraction $\phi_{\rm oil}\approx0.80$),
which is then diluted with TAP\,+\,Pluronic to the desired oil mass fraction.
Throughout, $\phi_{\rm oil}$ denotes the oil mass fraction;
since the density of castor oil is close to that of the aqueous phase
($\rho_{\rm oil}\simeq0.96$ g\,cm$^{-3}$),
mass and volume fractions differ by less than 2\% and can be compared directly to $\phi_{\rm c}$.
The protocol yields polydisperse droplets~\cite{Lestime2023}
with a mean radius $\langle R_{\rm drop}\rangle\simeq5~\mu$m,
comparable to the size of the algae
[see Fig.~\ref{fig:ActiveEmulsionSnapshot_and_OscillationMeasurement}(a)
and Fig.~\ref{supfig:Emulsion}],
and samples are gently rolled to ensure homogeneity prior to experiments.

The living inclusions are wild-type \textit{C.~reinhardtii} (CC125),
a puller-type biflagellated microswimmer.
Cultures are maintained on solid TAP and expanded in liquid TAP at $22^{\circ}$C
under a 16 h/8 h light--dark cycle.
Before mixing, cells are concentrated by centrifugation to reach the desired volume coverage
(typically $\sim60\%$ in confocal images of the continuous phase; see Fig.~\ref{supfig:coverage})
and gently dispersed in the mother emulsion.
Chlorophyll autofluorescence provides a direct marker of cell positions
within the droplet network [green in Fig.~\ref{fig:ActiveEmulsionSnapshot_and_OscillationMeasurement}(a)].

To ensure both emulsion stability and sustained algal motility,
we optimized the surfactant concentration by quantifying swimming dynamics
for $c_{\rm surf}=0$--50 CMC in the continuous phase (Fig.~\ref{supfig:MicroscopyTrackingPluronic}).
Characteristic velocities decrease only moderately between 0 and 10 CMC,
while motility is strongly reduced above.
A concentration of 10 CMC therefore provides a robust compromise,
yielding stable, biocompatible emulsions without significantly affecting cell activity.

\textit{Results.}--Confocal imaging confirms that the algae are homogeneously dispersed
within the droplet network [Fig.~\ref{fig:ActiveEmulsionSnapshot_and_OscillationMeasurement}(a)
and Supplementary Video S1].
Droplets form a dense, connected packing already for $\phi_{\rm oil}\gtrsim0.5$,
with the algae occupying interstitial pores.
This dense packing reflects the high Pluronic concentration required to stabilize the emulsion,
which induces an effective attraction between droplets through micellar depletion~\cite{Datta2011,Becu2006,Irani2014};
the continuous phase itself, with and without algae, displays only a weak elastic response
(Fig.~\ref{supfig:ContinuousPhase}).
Our emulsions are therefore attractive rather than purely repulsive packings,
and are expected to sustain a finite yield stress on both sides of RCP \cite{Datta2011}.

Rheological measurements are performed with a stress-controlled Anton Paar MCR 302 rheometer
in cone--plate geometry (cone angle $0.998^{\circ}$, truncation 0.101 mm, sand-blasted surfaces)
at $20^{\circ}$C, in a humidity chamber to minimize evaporation.
Samples are pre-sheared at $\dot{\gamma}=10~\mathrm{s^{-1}}$ for 30 s
and rested for 5 s before each run.
Oscillatory stress sweeps at $\omega=1$ Hz probe the viscoelastic response near yielding.
Yield stresses are extracted with the tangent method applied to $G'$ and $G''$~\cite{dinkgreve2016}:
$\sigma_y$ is defined as the stress at which the low-stress plateau of $G'$
intersects the high-stress decay regime (equivalently, where $G'$ and $G''$ cross
within the same interpolation scheme).
Each sample is measured at least three times and averaged over several emulsions
to ensure reproducibility.

Figure~\ref{fig:ActiveEmulsionSnapshot_and_OscillationMeasurement}(b)
shows a typical stress sweep at $\phi_{\rm oil}=0.62$.
Without algae, the emulsion exhibits a clear yielding transition:
$G'$ dominates at low stress and drops sharply near $\sigma_y$.
When motile algae are introduced, the crossover shifts to higher stress
and $\sigma_y$ increases by nearly a factor of two.
Motile inclusions in the continuous phase therefore reinforce the jammed droplet network,
increasing its resistance to yielding.

To disentangle biological activity from purely steric effects,
we repeated these measurements on emulsions containing algae immobilized
with Lugol's solution~\cite{Moreno2013},
which suppresses motility while leaving the cells intact.
The yield stress of the immobilized systems exceeds that of algae-free emulsions only slightly,
showing that the presence of inclusions alone provides at most marginal structural reinforcement;
it remains far below that of emulsions containing living, motile cells.
The dominant contribution to the enhancement of $\sigma_y$ therefore arises from active stresses
generated by the swimmers rather than from static geometric effects.

We now turn to the dependence on oil fraction.
As expected for jammed systems, both passive and active emulsions become increasingly rigid
with increasing $\phi_{\rm oil}$: droplets pack more densely, rearrangements are hindered,
and $\sigma_y$ grows accordingly [Fig.~\ref{fig:Results_ShearRheologyOscillation_All}(a,b)].
Across the whole range $\phi_{\rm oil}=0.40$--0.80, emulsions containing motile algae
display systematically higher yield stresses than both their immobilized and algae-free counterparts.
Since immobilized-algae emulsions are rheologically indistinguishable from algae-free ones
[Fig.~\ref{fig:ActiveEmulsionSnapshot_and_OscillationMeasurement}(b)],
we use the latter as the reference state in what follows:
they provide a more reproducible baseline, because dispersing algae--motile or immobilized--
introduces additional variability through fluctuations in cell concentration and local composition
(see Sec.~\ref{sec:emulsions} for details on sample preparation).

The absolute increase in yield stress between active and passive emulsions
grows from $\sim0.1$--0.5 Pa at low $\phi_{\rm oil}$ to nearly 100 Pa
at the highest oil fractions [Fig.~\ref{fig:Results_ShearRheologyOscillation_All}(c)].
Motile algae thus reinforce the passive droplet network over the entire range explored,
and increasingly so as the packing tightens.

\begin{figure*}[t]
\centering
\includegraphics{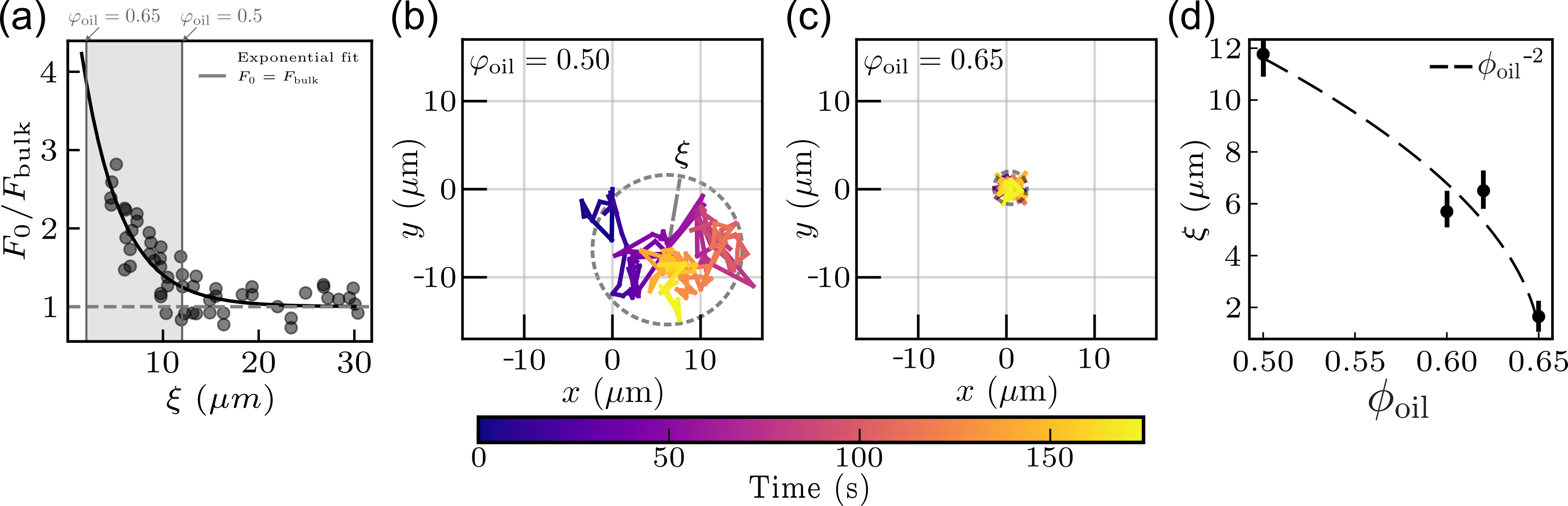}
\caption{\textbf{Confinement and active forces of algae in emulsions.}
(a) Active force exerted by \textit{C.~reinhardtii} as a function of confinement,
adapted from Ref.~\cite{Boddeker2020}.
The range of confinement explored in our emulsions is highlighted in grey;
these values are used to estimate the active stress in Eq.~(\ref{eq:sigma_phi})
and to compare it with the measured excess yield stress in
Fig.~\ref{fig:Results_ShearRheologyOscillation_All}(c).
(b),(c) Typical centre-of-mass trajectories of a motile alga in emulsions
at two oil fractions: (b) $\phi_{\rm oil}=0.65$ and (c) $\phi_{\rm oil}=0.50$,
recorded over 175 s.
(d) Characteristic confinement length $\xi$ as a function of oil fraction,
extracted from the mean-square displacement of trajectories such as those in (b),(c).
The dotted line is a power-law fit, $\xi\sim\phi_{\rm oil}^{-2}$,
used to propagate the confinement dependence of the active force shown in (a).}
\label{fig:MSD}
\end{figure*}

\textit{Mechanism.}--We interpret this reinforcement as an \emph{effective active depletion}
interaction between the passive, non-Brownian oil droplets, mediated by the motile algae.
Mixtures of passive and active particles are known to develop effective attractions
and to phase separate, because the active component continuously injects energy into the passive one
and generates stresses that mimic depletion forces~\cite{Wu2000,Angelani2011,Damman2024,Grober2023,Sinaasappel2026}.
In our emulsions, such activity-induced attractions strengthen the contact network
between droplets and hence raise the macroscopic yield stress.
Consistently, the effect is largest where the network is densest,
i.e. where each droplet has the largest number of contacts to reinforce.

A key ingredient of this mechanism is that the active forces generated by
\textit{C.~reinhardtii} are themselves amplified by confinement.
B\"oddeker \textit{et al.}~\cite{Boddeker2020}
measured propulsion forces increasing from $F^{\rm a}_{\rm bulk}\approx20$--30 pN in bulk
to $F^{\rm a}_{\rm conf}\approx90$ pN under strong confinement,
for gap sizes of a few micrometres [Fig.~\ref{fig:MSD}(a)].
In our emulsions, the droplet network naturally confines the algae in pores
whose characteristic size $\xi$ decreases as the oil fraction increases.

To quantify this confinement, we tracked individual algae by confocal microscopy
[Fig.~\ref{fig:MSD}(b,c)].
From the corresponding mean-square displacements we extracted the confinement length $\xi$
as the size of the region explored at long times (Fig.~\ref{supfig:MSDemulsion}).
We find that $\xi(\phi_{\rm oil})$ decreases continuously with oil fraction,
from slightly above one cell diameter at low $\phi_{\rm oil}$
to well below the cell size at the highest concentrations,
and is well described by $\xi\sim\phi_{\rm oil}^{-2}$ over the range explored
[Fig.~\ref{fig:MSD}(d)].

The active stress transmitted to the surrounding droplet network
can then be estimated as the force of a confined swimmer
acting over the cross-section of a droplet contact,
\begin{equation}
\sigma_{\phi}^{\rm a}\sim\frac{F^{\rm a}(\xi)}{\pi R_{\rm drop}^{2}},
\label{eq:sigma_a}
\end{equation}
where $F^{\rm a}(\xi)$ is the confinement-dependent propulsion force
of Ref.~\cite{Boddeker2020} and $R_{\rm drop}$ the mean droplet radius.
Over the relevant range, the measurements of B\"oddeker \textit{et al.}
are well captured by an exponential dependence on inverse confinement,
$F^{\rm a}(\xi)\simeq F_0\exp(\ell/\xi)$,
with $\ell$ a microscopic length of order the cell size.
Inserting the experimental scaling $\xi=\xi_0\,\phi_{\rm oil}^{-2}$
from Fig.~\ref{fig:MSD}(d) and expanding to leading order gives
\begin{equation}
\sigma_{\phi}^{\rm a}\simeq
\frac{F_0}{\pi R_{\rm drop}^{2}}
\left[1+\left(\frac{\phi_{\rm oil}}{\phi^{*}}\right)^{2}\right],
\qquad \phi^{*}=\sqrt{\xi_0/\ell},
\label{eq:sigma_phi}
\end{equation}
shown as the solid line in Fig.~\ref{fig:Results_ShearRheologyOscillation_All}(c).

This simple scaling predicts active stress increments
of $\sigma^{\rm a}_{\rm low}\approx1$ Pa at low oil fractions (weak confinement)
and $\sigma^{\rm a}_{\rm high}\approx30$ Pa at high oil fractions (strong confinement),
in good agreement with the measured increase in yield stress,
both in magnitude and in its evolution with $\phi_{\rm oil}$.
That an estimate built solely on independently measured swimmer forces
and on the independently measured confinement length reproduces the data
supports the interpretation that confinement-enhanced propulsion generates active stresses
large enough to reinforce the droplet contact network.

\textit{Conclusion.}--One of the grand challenges in soft matter and materials science
is to design materials whose mechanical properties can be programmed
through their microscopic constituents.
Our results demonstrate that biological activity provides such a route,
enhancing the strength and yield stress of emulsions beyond what passive inclusions achieve
and extending recent strategies based on the active structuring of complex fluids~\cite{Grober2023}.

By systematically comparing emulsions without algae, with immobilized algae,
and with motile \textit{Chlamydomonas reinhardtii},
and by combining rheology with single-cell tracking,
we showed that the increase in yield stress originates from the activity
of confined microswimmers rather than from their mere presence.
We propose that this reinforcement results from confinement-enhanced active forces,
which generate effective depletion-like attractions between neighbouring droplets
and strengthen the contact network.
More generally, our experiments reveal a feedback mechanism
in which the emulsion microstructure confines the swimmers,
confinement amplifies the forces they exert,
and these active stresses in turn reinforce the mechanical response of the emulsion.

Active emulsions thus emerge as a model system for studying jamming in active matter
and identify confinement as a powerful control parameter
coupling microscopic activity to macroscopic mechanics.
Our measurements also raise a broader question:
pusher-type swimmers fluidize simple suspensions~\cite{Sokolov2009,Lopez2015},
whereas the pullers used here rigidify a jammed one,
suggesting that the sign of the active contribution to the rheology
is set jointly by the swimming mechanism and by the state of the passive matrix.
Future work could explore higher swimmer concentrations,
activity controlled externally by light or chemical stimuli,
and other classes of microswimmers with distinct propulsion mechanisms,
such as pusher-type bacteria (\textit{e.g.}, \textit{E.~coli}).
Understanding how active stresses reorganize disordered contact networks
may provide general principles for engineering active complex fluids
with programmable mechanical properties.

\begin{acknowledgments}
We thank Annie Colin and Jeremie Palacci for interesting discussions and suggestions.
\end{acknowledgments}

\putbib
\end{bibunit}

\clearpage

\onecolumngrid

\setcounter{secnumdepth}{3}
\setcounter{section}{0}
\renewcommand{\thesection}{S\arabic{section}}
\renewcommand{\thesubsection}{\thesection.\arabic{subsection}}

\setcounter{figure}{0}
\setcounter{table}{0}
\setcounter{equation}{0}
\renewcommand{\thefigure}{S\arabic{figure}}
\renewcommand{\thetable}{S\arabic{table}}
\renewcommand{\theequation}{S\arabic{equation}}
\setcounter{secnumdepth}{3}
\setcounter{section}{0}
\renewcommand{\thesection}{S\arabic{section}}
\renewcommand{\thesubsection}{\thesection.\arabic{subsection}}
\makeatletter
\renewcommand{\p@subsection}{}
\makeatother

\begin{bibunit}

\begin{center}
{\large\bfseries Supplementary Material for\\[4pt]
``Active Reinforcement of Jammed Emulsions by Living Microswimmers''}\\[10pt]
Marie Corpart, Hugo Le Roux, Daniel Bonn, and Antoine Deblais\\[4pt]
\textit{Van der Waals-Zeeman Institute, Institute of Physics, University of Amsterdam,
1098 XH Amsterdam, Netherlands}
\end{center}
\vspace{6pt}

This Supplementary Material details the materials and the protocols used in this work: the formulation and preparation of the castor oil-in-water emulsions and their characterization (Sec.~\ref{sec:emulsions}), the culture, harvesting and immobilization of \textit{Chlamydomonas reinhardtii} and their incorporation into the emulsions (Sec.~\ref{sec:algae}), the imaging and single-cell tracking used to extract the confinement length $\xi$ (Sec.~\ref{sec:tracking}), and the rheological protocols and the determination of the yield stress (Sec.~\ref{sec:rheology}). It also reports the conversion between mass and volume oil fractions, the droplet size, the rheology of the continuous phases, and control measurements supporting the analysis presented in the main text.

Throughout this Supplementary Material we use the same notation as in the main text: $\phi_{\rm oil}$ denotes the \emph{mass} fraction of oil in the emulsion, $c_{\rm surf}$ the surfactant concentration expressed in units of the critical micellar concentration (CMC), $\xi$ the confinement length of the algae, and $\sigma_y$ the yield stress. Superscripts $^{\rm p}$ and $^{\rm a}$ refer to passive (algae-free) and active (motile algae) emulsions, respectively.

\section{Emulsions}
\label{sec:emulsions}

\subsection{Materials and formulation}

The emulsions used in this study are castor oil-in-water emulsions stabilized by the non-ionic triblock copolymer Pluronic F-127 (Sigma-Aldrich), a poloxamer of formula PEO$_{100}$PPO$_{65}$PEO$_{100}$ chosen for its biocompatibility with cell cultures. The dispersed phase is castor oil (Sigma-Aldrich), a triglyceride of density $\rho_{\rm oil}=\SI{0.961}{\gram\per\cubic\centi\metre}$ and viscosity $\eta_{\rm oil}\approx\SI{0.65}{\pascal\second}$ at \SI{20}{\celsius}. The continuous phase is TAP (Tris-Acetate-Phosphate, Gibco), the growth medium optimized for \textit{Chlamydomonas reinhardtii} cultures, so that the aqueous phase of the emulsion is also a viable environment for the algae. Nile Red (Sigma-Aldrich) is dissolved in the oil phase to render it fluorescent.

The CMC of Pluronic F-127 in TAP was measured to be \SI{3.3}{\gram\per\litre}. Unless stated otherwise, all emulsions are formulated with a continuous phase at $c_{\rm surf}=10$~CMC; the rationale for this choice is given in Sec.~\ref{sec:pluronic}.

\subsection{Emulsification protocol}

Emulsions are prepared in two steps: a concentrated mother emulsion is produced at high shear, and is subsequently diluted to the desired oil fraction. The mother emulsion is prepared as follows:

\begin{enumerate}[nolistsep]
    \item The continuous aqueous phase is prepared by dissolving Pluronic F-127 in TAP growth medium to reach 10~CMC (\SI{3.3}{\percent} by weight). The solution is left to homogenize for one day on a tube roller and then kept at \SI{4}{\celsius} until use.
    \item The dispersed phase is castor oil rendered fluorescent by dissolving $\sim\SI{0.3}{\percent}$ by weight of Nile Red.
    \item A first aliquot of the oil phase, equal in mass to the aqueous phase, is added and stirred with a Silverson L5M-A emulsifier at \SI{1000}{rpm} for \SI{1}{\minute}.
    \item The remaining oil is added continuously with a glass burette while the rotation speed is increased in steps of \SI{750}{rpm} every \SI{30}{\second}, reaching \SI{7000}{rpm} after \SI{4.5}{\minute}. The speed is then held constant for \SI{9.5}{\minute}.
    \item The speed is raised to \SI{8000}{rpm} for \SI{1}{\minute} and finally to \SI{9000}{rpm} for \SI{2}{\minute}.
    \item The emulsion is cooled in an ice bath for \SI{5}{\minute} to remove the heat generated during emulsification and to prevent thermal damage to the formulation. The internal mass fraction of the mother emulsion is $\phi_{\rm oil}=0.80$.
\end{enumerate}

Emulsions at lower oil fractions ($\phi_{\rm oil}=0.40$--$0.80$) are obtained by diluting the mother emulsion with the same aqueous phase at 10~CMC using a micropipette, so that the surfactant concentration in the continuous phase, and hence the strength of the depletion attraction between droplets, is identical in all samples. Diluted emulsions are briefly homogenized by hand with a cut pipette tip, which limits the shear applied to the droplet network, and then placed on a rolling device for at least \SI{5}{\minute} before use.

\subsection{Sample handling, stability and ageing}
\label{sec:ageing}

Bulk rheology requires at least \SI{20}{\milli\litre} of sample, which is also the volume above which ageing (creaming, sedimentation of the algae, coalescence~\cite{fredrick2010}) remains negligible over the duration of a measurement series. In practice, all emulsions are used within one hour of preparation and are returned to the rolling device between successive measurements. We verified that this protocol is sufficient by repeating a full stress sweep on the same sample at $t=0$ and $t=\SI{60}{\minute}$: the yield stress varies by less than \SI{5}{\percent} between the two runs. Emulsions containing algae are handled identically, with the additional constraint that samples are kept under ambient light and are used for at most two successive measurements after the cells have been dispersed, beyond which a measurable loss of motility is observed.

\subsection{Mass versus volume oil fraction}
\label{conversion}

Throughout the main text and this Supplementary Material, the oil content is expressed as a mass fraction $\phi_{\rm oil}$, which is the quantity directly controlled during preparation. Because the density of castor oil is close to that of the aqueous phase, the corresponding volume fraction
\begin{equation}
\phi_{\rm oil}^{\rm vol}=\frac{\phi_{\rm oil}/\rho_{\rm oil}}{\phi_{\rm oil}/\rho_{\rm oil}+(1-\phi_{\rm oil})/\rho_{\rm aq}}
\end{equation}
differs from $\phi_{\rm oil}$ by at most $0.01$ in absolute value over the whole range investigated (Tab.~\ref{suptab:conversion}), with $\rho_{\rm oil}=\SI{0.961}{\gram\per\cubic\centi\metre}$ and $\rho_{\rm aq}\simeq\SI{1.00}{\gram\per\cubic\centi\metre}$ for TAP\,+\,Pluronic. The distinction is therefore immaterial when comparing our data to the random close packing threshold $\phi_{\rm c}\approx0.64$, and mass and volume fractions are used interchangeably in that context.

\begin{table}[h]
    \centering
    \caption{\textbf{Conversion between the oil mass fraction $\phi_{\rm oil}$ used throughout this work and the corresponding volume fraction $\phi_{\rm oil}^{\rm vol}$}, computed with $\rho_{\rm oil}=\SI{0.961}{\gram\per\cubic\centi\metre}$ and $\rho_{\rm aq}=\SI{1.00}{\gram\per\cubic\centi\metre}$. The mass of dissolved surfactant is neglected, which affects $\phi_{\rm oil}^{\rm vol}$ by less than $0.002$.}
    \label{suptab:conversion}
    \begin{tabular}{cc}
    \toprule
    $\phi_{\rm oil}$ (mass) & $\phi_{\rm oil}^{\rm vol}$ (volume) \\
    \midrule
    0.40 & 0.410 \\
    0.45 & 0.460 \\
    0.50 & 0.510 \\
    0.55 & 0.560 \\
    0.60 & 0.610 \\
    0.62 & 0.629 \\
    0.65 & 0.659 \\
    0.70 & 0.708 \\
    0.75 & 0.757 \\
    0.80 & 0.806 \\
    \bottomrule
    \end{tabular}
\end{table}

\subsection{Microscopy and droplet size distribution}

Emulsions are imaged with an inverted confocal fluorescence microscope (Leica Microsystems, DMI4000~B) using a \SI{532}{\nano\metre} laser line to excite the Nile Red dissolved in the oil phase. Microscope slides and cover slips are separated by an imaging spacer (Grace Bio-Labs) of \SI{13}{\milli\metre} in diameter and \SI{0.12}{\milli\metre} in depth, which maintains a sufficient sample volume and minimizes evaporation during acquisition.

\begin{figure}[t]
    \centering
    \includegraphics[width=0.4\linewidth]{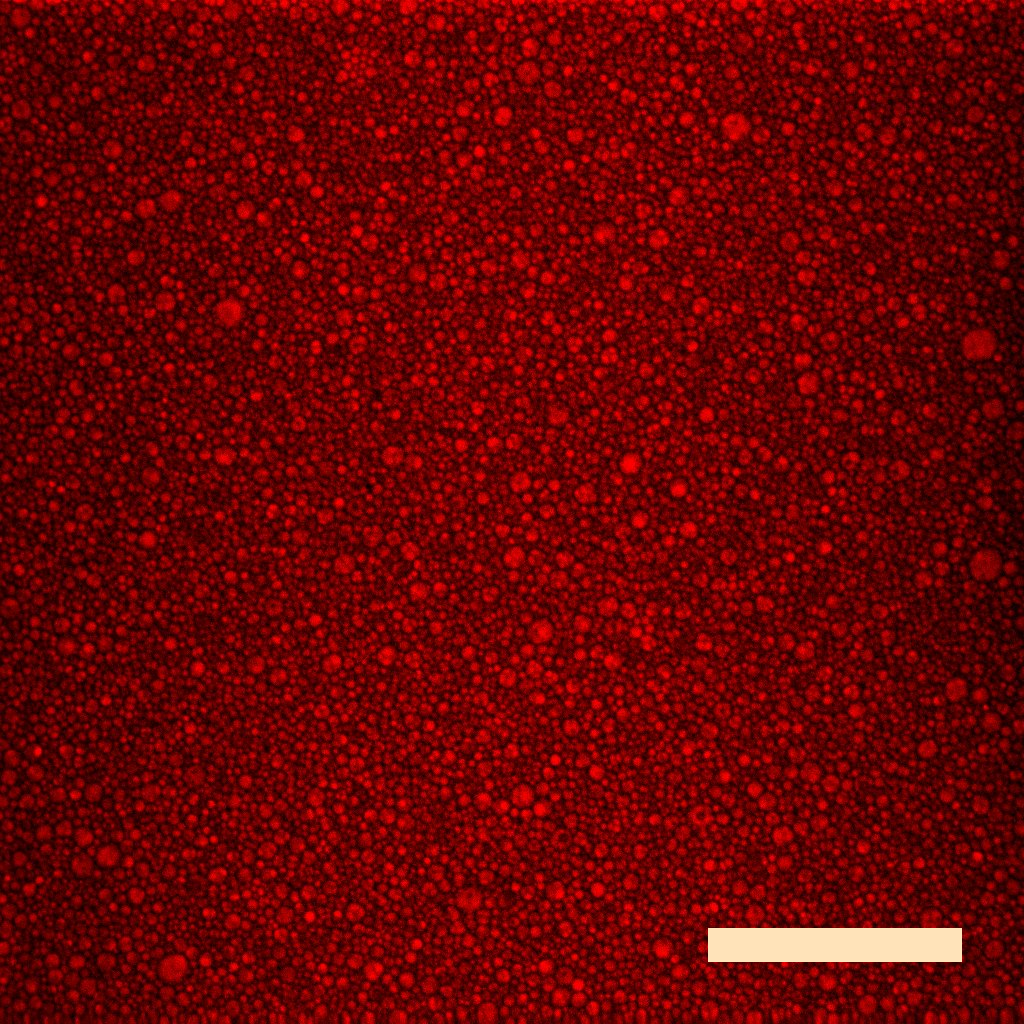}
    \caption{\textbf{Structure of the oil-in-water emulsions.} Confocal fluorescence image of an emulsion at $\phi_{\rm oil}=0.80$; the oil phase is labelled with Nile Red. Scale bar: \SI{100}{\micro\metre}.}
    \label{supfig:Emulsion}
\end{figure}

Droplet radii are extracted by manual segmentation of confocal slices [Fig.~\ref{supfig:Emulsion}] with the free software ImageJ, over typically $1000$ droplets. The emulsions are polydisperse, as expected for this emulsification method and as measured in previous studies~\cite{Lestime2023}, with a mean droplet radius $\langle R_{\rm drop}\rangle=\SI{5}{\micro\metre}$ and a polydispersity index of $0.36$~\cite{Lestime2023}. This mean radius is the value used in Eq.~(\ref{eq:sigma_a}) of the main text to convert the propulsion force of a confined alga into an active stress; it is also the relevant length to compare with the cell body diameter of \textit{C.~reinhardtii} ($\approx\SI{10}{\micro\metre}$).

\section{Algae}
\label{sec:algae}

\subsection{Strain and culture conditions}

We use the wild-type strain CC125 of \textit{Chlamydomonas reinhardtii}, a biflagellated, puller-type microalga with a cell body diameter of $\approx\SI{10}{\micro\metre}$ and two flagella of $\approx\SI{10}{\micro\metre}$ beating in a breaststroke fashion. Cultures are maintained on solid TAP agar plates and expanded in liquid TAP at \SI{22}{\celsius} under a \SI{16}{\hour}/\SI{8}{\hour} light--dark cycle. Cells are harvested in the exponential growth phase, when motility is highest.

\subsection{Harvesting and incorporation into the emulsions}

Before mixing, the culture is concentrated by centrifugation at $900\,g$ for \SI{20}{\minute}, a regime chosen to be gentle enough to preserve the flagella; the supernatant is removed and the pellet is resuspended in a small volume of TAP\,+\,Pluronic at 10~CMC. The concentrated suspension is then dispersed in the mother emulsion by gentle manual mixing, and the emulsion is diluted to the target oil fraction as described above. Because the algae are carried by the continuous phase, their concentration \emph{in that phase} is the quantity kept constant across samples; the number of cells per unit volume of emulsion therefore decreases as $\phi_{\rm oil}$ increases, and we account for this when comparing samples.

\begin{figure}[t]
    \centering
    \includegraphics[width=0.4\linewidth]{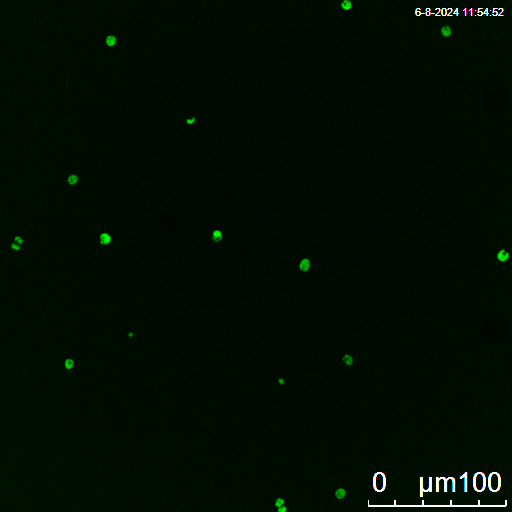}
    \caption{\textbf{Algae loading.} Typical confocal fluorescence image of an active emulsion showing the algae (green) dispersed in the continuous phase. The fluorescence signal is used to determine the initial volume concentration of algae; shown here is a slice of the continuous phase containing living algae, for a total volume of \SI{1.5}{\milli\litre}. Scale bar: \SI{100}{\micro\metre}.}
    \label{supfig:coverage}
\end{figure}

The resulting cell loading is quantified by the volume coverage measured on a slice of a confocal fluorescence image (Leica $10\times$ ACS APO objective, NA 0.3, pinhole \SI{94}{\micro\metre}, equivalent slice thickness \SI{11}{\micro\metre}). The centrifugation protocol described above yields a high volume coverage of algae in the mother continuous phase, typically $\sim\SI{60}{\percent}$ (Fig.~\ref{supfig:coverage}).

\subsection{Immobilization with Lugol's solution}
\label{sec:lugol}

To separate the contribution of activity from that of the mere presence of the inclusions, a set of emulsions is prepared with algae whose motility has been suppressed with Lugol's solution~\cite{Moreno2013}, a standard fixative for phytoplankton that arrests the flagellar beating while preserving the cell body and its size. Lugol is added to the concentrated cell suspension at a final concentration of \SI{0.1}{\percent} by weight and left to act for \SI{10}{\minute} before the suspension is incorporated into the emulsion, following exactly the protocol used for motile cells. We verified the absence of residual motion by tracking the immobilized cells over \SI{100}{\second}, which yields a mean-square displacement indistinguishable from zero.

Two controls are important here. First, Lugol itself does not modify the mechanics of the continuous phase: the rheology of TAP\,+\,Pluronic with and without Lugol is indistinguishable (Fig.~\ref{supfig:ContinuousPhase}). Second, immobilized cells have the same size as motile ones, so that the steric contribution of the inclusions is unchanged. Together these controls justify the statement made in the main text that the difference between immobilized and motile samples isolates the effect of activity.

\section{Imaging and tracking of the algae}
\label{sec:tracking}

\subsection{Acquisition}

The dynamics of the algae in their culture medium are followed by bright-field microscopy (Leica Microsystems, DMI4000~B) with a Basler camera (acA2040-120uc) at \SI{20}{\hertz}. Inside the emulsions, bright-field imaging is not usable because the droplet network scatters light strongly; the cells are therefore imaged by their chlorophyll autofluorescence on the same inverted confocal microscope, excited at \SI{635}{\nano\metre} and acquired at \SI{10}{\hertz} with a Leica $10\times$ ACS APO objective (NA 0.3) and a pinhole of \SI{94}{\micro\metre}. Acquisitions in emulsions last \SI{200}{\second}, long enough to resolve the plateau of the mean-square displacement at the highest oil fractions.

\subsection{Tracking and mean-square displacements}

Trajectories are reconstructed with a Python~3.12.7 script based on the \texttt{trackpy} module~\cite{Crocker1996}, and the mean-square displacement is computed as
\begin{equation}
\mathrm{MSD}(\tau)=\left\langle\left\lvert\boldsymbol{r}(t+\tau)-\boldsymbol{r}(t)\right\rvert^{2}\right\rangle,
\end{equation}
where the average runs over time origins $t$ and over trajectories. Before averaging, we discard cells moving slower than two cell diameters per second and trajectories shorter than \SI{50}{\second}. This filtering removes non-motile or dead cells and poorly resolved tracks, so that the ensemble-averaged MSD is representative of the active motion.

In the culture medium, the MSD displays the two regimes expected for a persistent random walker: a ballistic regime $\langle\Delta r^{2}\rangle=v_{0}^{2}\tau^{2}$ at short times and a diffusive regime $\langle\Delta r^{2}\rangle=4D\tau$ at long times, separated by a characteristic time $\tau_{\rm c}$ [Fig.~\ref{supfig:MicroscopyTrackingPluronic}(b)]. We measure $\tau_{\rm c}=\SI{1.47}{\second}$, $v_{0}=\SI{28.2}{\micro\metre\per\second}$ and $D=3.7\times10^{2}~\si{\micro\metre\squared\per\second}$. These three quantities are not independent: matching the two regimes at $\tau=\tau_{\rm c}$ gives $D=v_{0}^{2}\tau_{\rm c}/4\simeq2.9\times10^{2}~\si{\micro\metre\squared\per\second}$, in good agreement with the measured value and confirming the consistency of the analysis. The persistence length of a free cell, $v_{0}\tau_{\rm c}\simeq\SI{41}{\micro\metre}$, is about four cell diameters; it sets the scale against which the confinement imposed by the droplet network must be compared.

\subsection{Effect of the Pluronic concentration on motility}
\label{sec:pluronic}

\begin{figure*}[t]
    \centering
    \includegraphics[width=1\linewidth]{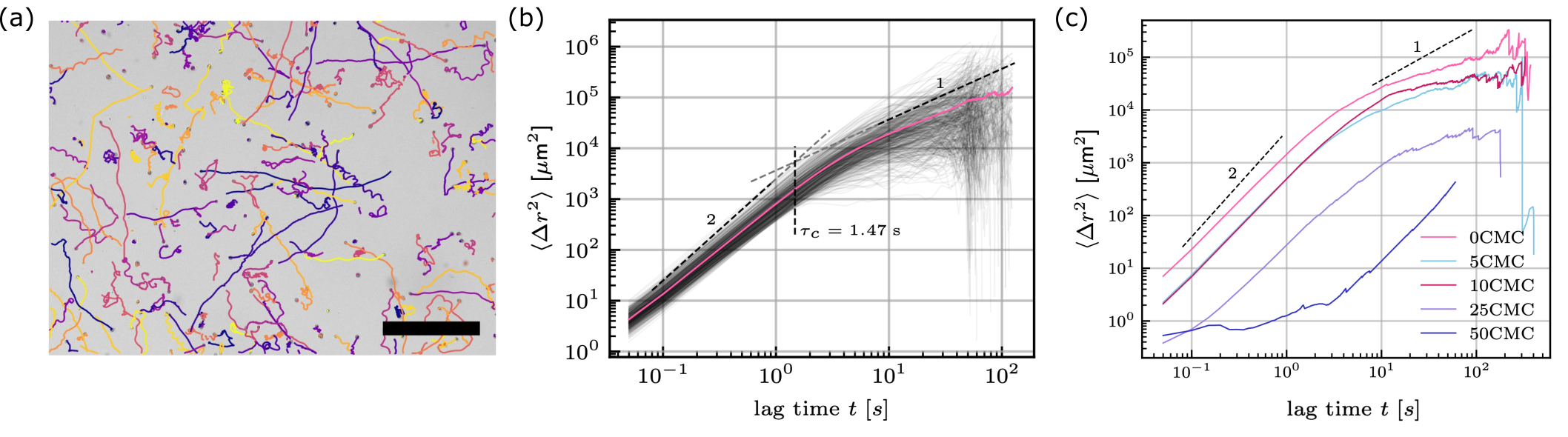}
    \caption{\textbf{Microscopy and tracking of CC125 algae in TAP.} (a)~Bright-field image of the algae in TAP with the reconstructed tracks superimposed. Scale bar: \SI{200}{\micro\metre}. (b)~Individual MSDs recorded over \SI{10}{\minute} (black) and ensemble-averaged MSD (pink). $\tau_{\rm c}$ denotes the characteristic time separating the ballistic and diffusive regimes. (c)~Ensemble-averaged MSDs of algae suspended in TAP at Pluronic concentrations $c_{\rm surf}=0$--50~CMC.}
    \label{supfig:MicroscopyTrackingPluronic}
\end{figure*}

The surfactant concentration was chosen as high as possible, to maximize emulsion stability and the depletion attraction between droplets, while preserving a sufficient level of algal activity. We therefore measured the swimming dynamics in TAP containing Pluronic F-127 at $c_{\rm surf}=0$, 5, 10, 25 and 50~CMC [Fig.~\ref{supfig:MicroscopyTrackingPluronic}(c)]. Applying the filtering described above, the characteristic time increases moderately from $\tau_{\rm c}=\SI{1.47}{\second}$ at 0~CMC to \SI{1.97}{\second} at 5~CMC and \SI{2.15}{\second} at 10~CMC. Over the same range the initial speed decreases from $v_{0}=\SI{28.2}{\micro\metre\per\second}$ at 0~CMC to \SI{22.4}{} and \SI{22.3}{\micro\metre\per\second} at 5 and 10~CMC, and then collapses to \SI{5.5}{} and \SI{1.22}{\micro\metre\per\second} at 25 and 50~CMC, where the cells are essentially immotile.

We conclude that the behaviour of the algae is only weakly affected up to 10~CMC, while higher concentrations alter it substantially. A concentration of 10~CMC is therefore a reasonable compromise between emulsion stability, favoured by high surfactant content, and cell motility, favoured by low surfactant content; it is used for all the emulsions reported in the main text.

\subsection{Dynamics inside the emulsions and extraction of the confinement length}
\label{sec:confinement}

\begin{figure*}[t]
    \centering
    \includegraphics[width=0.8\linewidth]{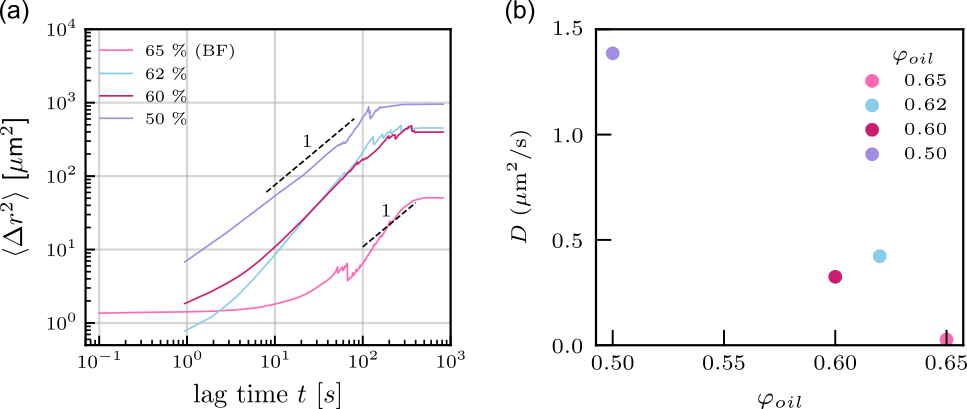}
    \caption{\textbf{Mean-square displacements of algae confined in the emulsions.} (a)~Ensemble-averaged MSDs of the algae for oil fractions $\phi_{\rm oil}=0.50$--$0.65$, showing the progressive suppression of the long-time transport and the emergence of a plateau. (b)~Diffusion coefficients extracted from the long-time behaviour of the MSDs, for the oil fractions accessible to tracking.}
    \label{supfig:MSDemulsion}
\end{figure*}

Inside the emulsions the picture changes qualitatively: the algae no longer explore space freely but remain trapped in the interstitial pores of the droplet network, exploring a region of finite extent before being able to squeeze into a neighbouring pore [see the trajectories in Fig.~\ref{fig:MSD}(b,c) of the main text]. The MSD accordingly departs from the free-swimming behaviour and saturates at long lag times [Fig.~\ref{supfig:MSDemulsion}].

We define the confinement length as $\xi=\sqrt{\mathrm{MSD}_{\rm plateau}}$, the size of the region explored at long times. $\xi$ decreases monotonically with the oil fraction (see main text), from \SI{12}{\micro\metre} at $\phi_{\rm oil}=0.50$, slightly larger than a cell diameter, down to \SI{2}{\micro\metre} at $\phi_{\rm oil}=0.65$, well below the cell size, so that the cells are then squeezed between neighbouring droplets. Over this range the data are described by a power law $\xi\sim\phi_{\rm oil}^{-2}$ [Fig.~\ref{fig:MSD}(d) of the main text]. This empirical scaling is used in the main text only to propagate the measured confinement into the force--confinement relation of Ref.~\cite{Boddeker2020}, and no physical meaning is attached to the exponent itself.

\section{Rheology}
\label{sec:rheology}

\subsection{Setup and measurement protocol}

Rheological measurements are performed with a stress-controlled rheometer (Anton Paar, MCR~302) in a cone-plate geometry (Anton Paar CP50-1/S, cone angle \SI{0.998}{\degree}, truncation \SI{0.101}{\milli\metre}) with sand-blasted surfaces to suppress wall slip, at \SI{20}{\celsius}. All measurements are carried out in a humidity chamber to limit evaporation, which is critical for the samples containing living cells. Before each run the sample is pre-sheared at $\dot\gamma=\SI{10}{\per\second}$ for \SI{30}{\second} and then left to rest for \SI{5}{\second}, so that every measurement starts from a reproducible, rejuvenated state.

Oscillatory stress sweeps are performed at a fixed frequency $\omega=\SI{1}{\hertz}$ over two decades of stress centred on the yield stress of the sample; this frequency lies in the plateau region of the loss modulus and keeps the total measurement time short enough to avoid ageing (Sec.~\ref{sec:ageing}). Each sample is measured at least three times, and between one and five independently prepared emulsions are used per condition; the error bars reported in the main text are standard errors of the mean over these repeats.

\subsection{Determination of the yield stress}

The yield stress reported in the main text is extracted from the oscillatory stress sweeps using the tangent method~\cite{dinkgreve2016}: $\sigma_y$ is defined as the stress at which the low-stress plateau of $G'$ intersects the high-stress decay of $G'$, the intersection being obtained by interpolation with the \texttt{scipy} module in Python~3.12.7.

\subsection{Rheology of the continuous phases}

\begin{figure*}[t]
    \centering
    \includegraphics[width=0.8\linewidth]{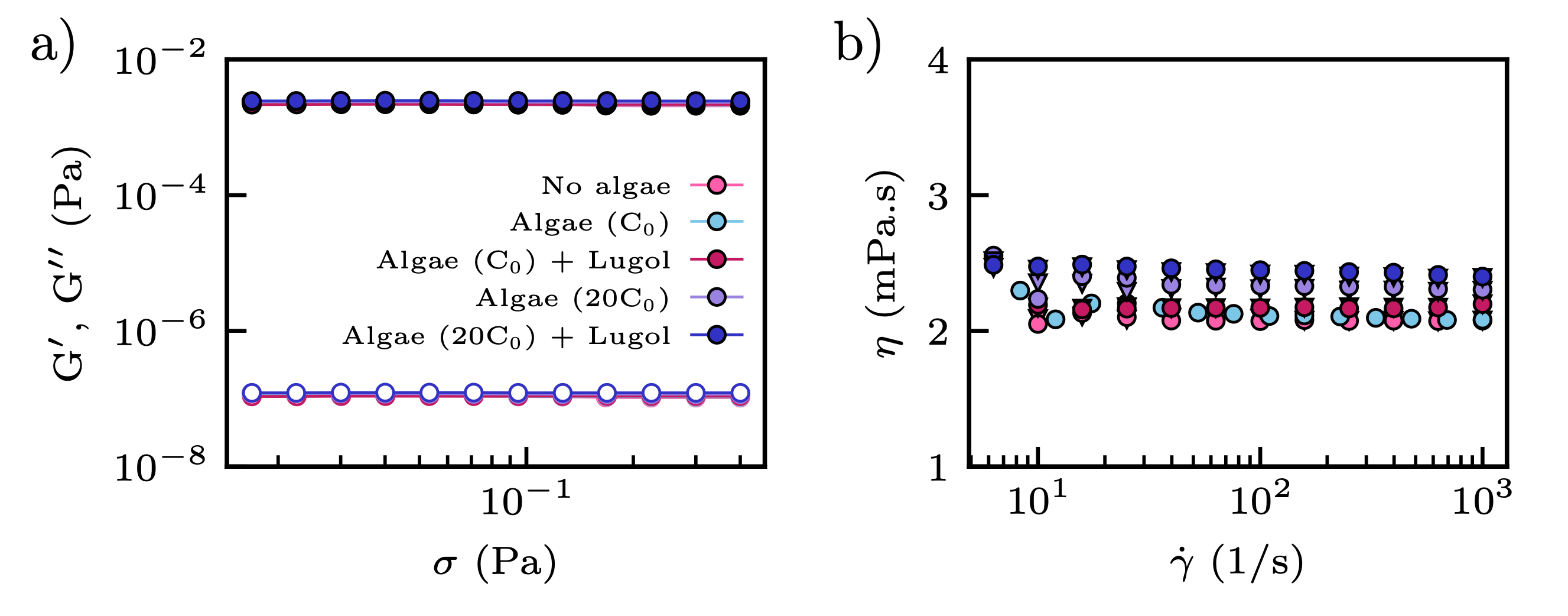}
    \caption{\textbf{Rheology of the continuous phases.} (a)~Storage $G'$ (open symbols) and loss $G''$ (filled symbols) moduli of the different continuous phases as a function of the applied shear stress. (b)~Viscosity of the same continuous phases as a function of shear rate.}
    \label{supfig:ContinuousPhase}
\end{figure*}

To confirm that the reinforcement reported in the main text originates from the droplet network and not from the suspending fluid, we characterized the continuous phases on their own: TAP, TAP\,+\,Pluronic at 10~CMC, and the same solution containing algae at the concentration used in the emulsions, motile and immobilized. All of them display only a weak elastic response and a viscosity close to that of water (Fig.~\ref{supfig:ContinuousPhase}), several orders of magnitude below the moduli measured on the emulsions. The active contribution to $\sigma_y$ therefore cannot be attributed to a change in the rheology of the continuous phase.

\section{Supplementary video}

\noindent\textbf{Video S1.} Confocal reconstruction of an active emulsion at $\phi_{\rm oil}=0.50$, showing the oil droplets (Nile Red, red) and the algae (chlorophyll autofluorescence, green) dispersed in the interstitial pores. Duration \SI{60}{\second}, acquisition rate \SI{0.15}{fps}, scale bar \SI{100}{\micro\metre}.

\putbib
\end{bibunit}

\end{document}